\documentstyle{article}
\def\be{\begin{equation}}
\def\ee{\end{equation}}
\def\ba{\begin{eqnarray}}
\def\ea{\end{eqnarray}}
\begin{document}
\newtheorem{th}{Donotwrite}[section]
\newtheorem{definition}[th]{Definition}
\newtheorem{theorem}[th]{Theorem}
\newtheorem{prop}[th]{Proposition}
\newtheorem{lemma}[th]{Lemma}
\newtheorem{corollary}[th]{Corollary}
\newtheorem{remark}[th]{Remark}
\newtheorem{example}[th]{Example}
\newtheorem{conjecture}[th]{Conjecture}
\def\qed{~\rule{1mm}{2.5mm}}
\def\dfrac#1#2{{\displaystyle\frac{#1}{#2}}}
\title{Discrete dynamical systems with 
$W(A^{(1)}_{m-1} \times A^{(1)}_{n-1})$ symmetry}
\author{
Kenji KAJIWARA$^a$, Masatoshi NOUMI$^b$ and Yasuhiko YAMADA$^b$ \\
\\
{\normalsize a: Department of Electrical Engineering, Doshisha University}\\
{\normalsize b: Department of Mathematics, Kobe University}
}
\maketitle
\date

\begin{abstract}
We give a birational realization of affine Weyl group of type
$A^{(1)}_{m-1} \times A^{(1)}_{n-1}$.  We apply this representation to
construct some discrete integrable systems and discrete Painlev\'e
equations.  Our construction has a combinatorial counterpart through the
ultra-discretization procedure.
\end{abstract}

\section{Introduction}

It is known that continuous and discrete Painlev\'e equations admit
affine Weyl group symmetry. These symmetries are realized as birational
B\"acklund transformations and can be described in universal way for
general root systems.\cite{NY2} On the other hand, in recent studies on
(ultra-)discrete integrable systems, certain affine Weyl group
representation has been obtained.\cite{Y} This representation is
``tropical'' (i.e. given in terms of 
subtraction-free birational mapping) and has a
combinatorial counterpart through the ultra-discretization procedure.
The aim of this paper is to reveal the relation between these two affine
Weyl group representations. In fact, we show that they are essentially
the same. 

In section 2, we recall the definition of birational transformations
$\pi, r_0, \ldots, r_m$ and $\rho,s_0, \ldots, s_n$ acting on $mn$ variables
$x_{ij}$ ($i=1,\ldots,m$, $j=1,\ldots,n$). The main result
(Theorem 1.1) states that these transformations give a representation of
affine Weyl group of type $A^{(1)}_{m-1} \times A^{(1)}_{n-1}$.
We give the proof of Theorem 1.1 in section 3.
In section 4, we describe the discrete dynamical systems arising
from our affine Weyl group representation. 
The systems can be viewed as a version of 
discrete Toda equation and its generalizations.
We also discuss the relation with $q$-Painlev\'e equation studied in \cite{KNY}.

\section{Affine Weyl group actions}

Let $m$, $n$ be positive integers and
$K={\bf C}(x)$ be the field of rational functions on $mn$ variables
$x_{ij}$, $i=1,2,\ldots,m$, $j=1,2,\ldots,n$.
We extend the indices $i,j$ of $x_{ij}$ for $i,j \in {\bf Z}$
by the condition $x_{i+m,j}=x_{i,j+n}=x_{ij}$.

Define algebra automorphisms 
$\pi$, $\rho$, $r_i$ and $s_j$ ($i \in {\bf Z}/m {\bf Z}$,
$j \in {\bf Z}/n {\bf Z}$) on the field $K$
as follows:
\ba
&&
\pi(x_{ij})=x_{i+1,j}, \quad \rho(x_{i,j})=x_{i,j+1}, \\[1mm]
&&r_i(x_{ij})=x_{i+1,j} 
\frac{P_{i,j-1}}{P_{ij}}, \quad
r_i(x_{i+1,j})=x_{ij}
\frac{P_{ij}}{P_{i,j-1}}, \\[1mm]
&&r_k(x_{ij})=x_{ij}, \quad (k \neq i,i+1),\\[1mm]
&&s_j(x_{ij})=x_{i,j+1} 
\frac{Q_{i-1,j}}{Q_{ij}}, \quad
s_j(x_{i,j+1})=x_{ij}
\frac{Q_{ij}}{Q_{i-1,j}}, \\[1mm]
&&s_k(x_{ij})=x_{ij}, \quad (k \neq j,j+1),
\ea 
where 
\ba
&&P_{ij}=\sum_{a=1}^{n} \left( 
\prod_{k=1}^{a-1} x_{i,j+k} 
\prod_{k=a+1}^{n} x_{i+1,j+k} \right), \\[2mm]
&&Q_{ij}=\sum_{a=1}^{m} \left( 
\prod_{k=1}^{a-1} x_{i+k,j} 
\prod_{k=a+1}^{m} x_{i+k,j+1} \right).
\ea

The following is the main result of this note.
\begin{theorem} \label{th:main}
$\langle \pi, r_0,r_1,\cdots,r_{m-1} \rangle$
and $\langle \rho, s_0,s_1,\cdots,s_{n-1} \rangle$
generate the extended affine Weyl group 
$\widetilde{W}(A^{(1)}_{m-1})$ and
$\widetilde{W}(A^{(1)}_{n-1})$.
Moreover these two actions 
$\widetilde{W}(A^{(1)}_{m-1})$ and
$\widetilde{W}(A^{(1)}_{n-1})$ mutually commute.
\end{theorem}

In our previous work \cite{Y}, the first part of this theorem has been proved 
and the second part was conjectured. 
We will prove Theorem \ref{th:main} in the next section.

\begin{example}{\rm
For the case of $(m,n)=(2,3)$. Let us write the $x_{ij}$
variables as
\be
X=\left[\begin{array}{ccc}
x_{11}&x_{12}&x_{13}\\
x_{21}&x_{22}&x_{23}
\end{array}\right]
=\left[\begin{array}{ccc}
x_1&x_2&x_3\\
y_1&y_2&y_3
\end{array}\right]
\ee
Then the actions $r_1, r_0$, $\pi$, $s_1, s_2, s_0$ and $\rho$
are given as follows.
\be
\begin{array}l
\displaystyle
r_1(x_1)=y_1\frac{x_1x_2+x_1y_3+y_2y_3}{x_2x_3+x_2y_1+y_3y_1},\quad
r_1(y_1)=x_1\frac{x_2x_3+x_2y_1+y_3y_1}{x_1x_2+x_1y_3+y_2y_3},\\[5mm]
\displaystyle
r_1(x_2)=y_2\frac{x_2x_3+x_2y_1+y_3y_1}{x_3x_1+x_2y_2+y_1y_2},\quad
r_1(y_2)=x_2\frac{x_3x_1+x_2y_2+y_1y_2}{x_2x_3+x_2y_1+y_3y_1},\\[5mm]
\displaystyle
r_1(x_3)=y_3\frac{x_3x_1+x_2y_2+y_1y_2}{x_1x_2+x_1y_3+y_2y_3},\quad
r_1(y_3)=x_3\frac{x_1x_2+x_1y_3+y_2y_3}{x_3x_1+x_2y_2+y_1y_2},
\end{array}
\ee
$\pi(x_i)=y_i$, $\pi(y_i)=x_i$, $r_0=\pi r_1 \pi$  and
\be
\begin{array}{lll}
\displaystyle
s_1(x_1)=x_2 \frac{x_1+y_2}{x_2+y_1},&
\displaystyle
s_1(x_2)=x_1 \frac{x_2+y_1}{x_1+y_2},&
s_1(x_3)=x_3,\\[5mm]
\displaystyle
s_1(y_1)=y_2 \frac{x_2+y_1}{x_1+y_2},&
\displaystyle
s_1(y_2)=y_1 \frac{x_1+y_2}{x_2+y_1},&
s_1(y_3)=y_3,
\end{array}
\ee
$\rho(x_i)=x_{i+1}$,
$\rho(y_i)=y_{i+1}$ ,$ i \in {\bf Z}/(3 {\bf Z})$, 
$s_2=\rho s_1 \rho^{-1}$ and $s_0=\rho^2 s_1 \rho^{-2}$.
} 
\end{example}

\section{Proof of Theorem \ref{th:main}}

{\bf The first part} (affine Weyl group relations): 
Though this part of the theorem has already been proved in \cite{Y} 
by direct computation, we give more conceptual proof here.
We prove that $s_i$ ($i=1,\cdots,n-1$) satisfy the same relations as 
those for the adjacent permutations $\sigma_i=(i,i+1) \in S_{n}$.
The relations containing $s_0$ follow from the cyclic symmetry
of the definition (4,5,7).

Using the identities,
\be
x_{i,l+1}Q_{i-1,l}-x_{il}Q_{il}=k_{l+1}-k_l, \quad
k_l=\prod_{i=1}^m x_{il}.
\ee
we see that the automorphisms $s_j$ satisfy the relations,
\be\label{eq:xyrel}
x_i y_i=x'_iy'_i, \quad x_i+y_{i+1}=x'_i+y'_{i+1},
\ee
where $x_i=x_{ij}$, $y_i=x_{i,j+1}$,
$x'_i=s_j(x_{ij})$ and $y'_i=s_j(x_{i,j+1})$.
The relations (\ref{eq:xyrel}) can be written in matrix form
as
\be\label{eq:AAAA}
A(x) A(y)=A(x') A(y'),
\ee
where
\be
A(x)=\left(
\begin{array}{ccccc}
x_1&1& \cr
&x_2&1 \cr
&&\ddots&\ddots \cr
&&&x_{m-1}&1 \cr
z&&&&x_m
\end{array}
\right).
\ee
The relations (\ref{eq:xyrel}) essentially characterize the transformation
 $s_j$.
More generally we have the following,
\begin{lemma}\label{lem:soln}
For given $mn$ variables $x_j=(x_{1j},x_{2j},\cdots,x_{mj})$ and
$mn$ unknowns $x'_j=(x'_{1j},x'_{2j},\cdots,x'_{mj})$,
a system of algebraic equations
\be\label{eq:AAAn}
A(x_1)A(x_2)\cdots A(x_n)=A(x'_1)A(x'_2)\cdots A(x'_n),
\ee
has $n!$ solutions.
Each of the solution corresponds to a permutation $\sigma \in S_n$
and characterized by additional condition
\be\label{eq:sigmacons}
x'_{1j}x'_{2j}\cdots x'_{mj}=
x_{1 \sigma(j)}x_{2 \sigma(j)}\cdots x_{m \sigma(j)}.
\ee
\end{lemma}

\noindent
{\it Proof of Lemma \ref{lem:soln}}.\quad
>From the definition of the transformation $s_j$  (4,5,7),
it follows that
\be
A(s_j(x_1))\cdots A(s_j(x_n))=A(x_1)\cdots A(x_n),
\ee
and
\be
s_j(x_{1k}\cdots x_{mk})=x_{1\sigma_j(k)}\cdots x_{m\sigma_j(k)},
\quad k=1,\cdots, n,
\ee
where $\sigma_j=(j,j+1)$.
This implies that for any permutation $\sigma$  given as a product
$\sigma=\sigma_{i_1}\cdots \sigma_{i_k}$,
\be
x'_{ij}=s_{i_1}\cdots s_{i_k}(x_{ij}),
\ee
is a solution of (\ref{eq:AAAn}) and (\ref{eq:sigmacons}). 
Hence we see that there exist at least $n!$ solutions for 
(\ref{eq:AAAn}).

Let us next show that such solution is unique, namely we prove that the
number of solutions for (\ref{eq:AAAn}) is at most $n!$.
We discuss the case $n=3$ for simplicity.
Consider a system of algebraic equations for $3m$ unknown variables
$(x'_i,y'_i,z'_i)$ given by
\be\label{eq:AAA3}
A(x)A(y)A(z)=A(x')A(y')A(z'),
\ee
or
\be
a_i=a_i', \quad b_i=b_i', \quad c_i=c_i',
\ee
where
\be
a_i=x_i+y_{i+1}+z_{i+2}, \quad
b_i=x_i y_i+x_i z_{i+1}+y_{i+1}z_{i+1}, \quad
c_i=x_iy_iz_i;
\ee
$a_i',b_i'$ and $c_i'$ are defined similarly.
By simple elimination of variables $y'_i$ and $z'_i$, 
one has algebraic equations for $x'_i$
\be
x'_{i-1}x'_ix'_{i+1}-a_{i-1}x'_ix'_{i+1}+b_i x'_{i+1}-c_{i+1}=0,
\ee
or equivalently
\be\label{eq:Rrec}
x'_{i+1} \psi_i=R_{i+1} \psi_{i+1}, \quad
R_{i+1}=
\left[
\begin{array}{ccc}
&1\\&&1\\c_{i+1}&-b_{i}&a_{i-1}
\end{array}
\right], \quad
\psi_{i}=
\left[
\begin{array}{c}
1\\x'_{i}\\x'_{i-1}x'_{i}
\end{array}
\right].
\ee
Due to the periodicity $x'_{i+m}=x'_i$, we have
\be
(x'_1x'_2\cdots x'_{m}) \psi_1=(R_{m}\cdots R_{2}R_{1}) \psi_1.
\ee
Hence, $\psi_1$ is one of the three eigenvectors of 
$R_{m}\cdots R_{2}R_{1}$.
Once $\psi_1$ is determined other $\psi_i$'s and $x'_i$'s are 
determined uniquely by using (\ref{eq:Rrec})  repeatedly. 
Then we have the equation for $y'_i$'s and $z_i'$'s 
\be
A(x')^{-1}A(x)A(y)A(z)=A(y')A(z').
\ee
By similar argument we see that this equation has two solutions.
Hence we have at most $3!$ solutions for (\ref{eq:AAA3}).\qed

If $\sigma=\sigma_{i_1}\cdots\sigma_{i_k}=
\sigma_{j_1}\cdots \sigma_{j_l}$ then 
\be
s_{i_1}\cdots s_{i_k}(x_{1r}\cdots x_{mr})=
x_{1\sigma(r)}\cdots x_{m\sigma(r)}=
s_{j_1}\cdots s_{j_l}(x_{1r}\cdots x_{mr}),
\ee
and we have $s_{i_1}\cdots s_{i_k}=s_{j_1}\cdots s_{j_l}$
due to Lemma \ref{lem:soln}.
This means that all the relations satisfied by $\sigma_i$'s 
also hold for $s_i$'s.

\medskip
\noindent
{\bf The second part} (commutativity): 
Let us put $G_i(u)=1+u E_{i+1,i}$ for $i \in [1,m-1]$ and
$G_0(u)=1+{\displaystyle\frac{u}{z}}E_{1,m}$. We also put
\be
M=A(x_1)A(x_2)\cdots A(x_n),
\ee
where $x_j=(x_{1j}, \cdots x_{mj})$. 
We remark that
Lemma \ref{lem:soln}  also implies the invariance of 
$M$ under the actions of $s_j$ ($j=1,2,\ldots,n-1$).
The action of $r_k$ on $M$ is described as follows. 
\begin{lemma}\label{lem:GM}
\be\label{eq:GM}
G_k(u)M=r_k(M)G_k(u),
\qquad
u=\frac{h_k-h_{k+1}}{P_{k,0}}. 
\ee
\end{lemma}
\noindent
{\it Proof.} \quad
Putting 
\be
u_j=\frac{h_k-h_{k+1}}{P_{kj}}\qquad(j=0,1,\ldots,n),
\ee
the action of $r_k$ can be rewritten in the form
\be
\begin{array}{c}\smallskip
r_k(x_{kj})=x_{kj}-u_j,\quad
r_k(x_{k+1,j})=x_{k+1,j}+u_{j-1},
\quad\\
r_k(x_{ij})=x_{ij}\ \quad (i\ne k,k+1),
\end{array}
\ee
where we have used the identity
\be
x_{k+1,j}P_{k,j-1}-x_{kj}P_{kj}=h_{k+1}-h_k, \quad
h_k=\prod_{j=1}^n x_{kj}.
\ee
The relations above are equivalent to the matrix equation
\be
G_k(u_{j-1})A(x_j)=A(r_k(x_j))G_k(u_j)
\quad(j=1,\ldots,n). 
\ee
Hence we have
\be
G_k(u_0)A(x_1)\cdots A(x_n)=
A(r_k(x_1))\cdots A(r_k(x_n))G_k(u_n).  
\ee
Due to the periodicity $u_0=u_n(=u)$,  we have
\be
G_k(u) M=r_k(M) G_k(u),\quad u=\frac{h_k-h_{k+1}}{P_{k,0}},
\ee
where $M=A(x_1)\cdots A(x_n)$. \qed
\par\medskip\noindent
We remark that the parameter $u$ can also be determined 
from $M$ by the formula 
\be\label{eq:GM2}
u=\left. \frac{M_{kk}-M_{k+1,k+1}}{M_{k,k+1}}\right\vert _{z=0}. 
\ee
\par\medskip
The commutativity of the actions of the two Weyl groups
in Theorem \ref{th:main}
follows from Lemma \ref{lem:GM}, 
because the matrix $M$ is invariant under the actions of $s_j$
($j=1,2,\ldots,n-1$). The commutativity $r_i s_0=s_0 r_i$ follows
from $r_0=\rho^{-1} s_1 \rho$ and $r_i \rho=\rho r_i$.
Thus, the proof of Theorem \ref{th:main} is completed.
\par\medskip
In this proof, we have treated the actions $s_i$ and $r_j$ asymmetrically.
However the argument can be applicable for both of them, 
because the roles of $s_i$ and $r_j$ can be exchanged 
with each other by the transposition of the matrix $X=(x_{ij})$.

\section{Discrete dynamical systems}

The affine Weyl group 
$\widetilde{W}(A^{(1)}_{m-1}) \times \widetilde{W}(A^{(1)}_{n-1})$
has a translation subgroup ${\bf Z}^{m-1} \times {\bf Z}^{n-1}$.
If we consider a part of these translations as providing a 
discrete dynamical system, the commutant of them 
can be viewed as its B\"acklund transformations. 

\par\medskip
\begin{example} \label{ex:m2}\quad {\rm
When $m=2$, the translation subgroup of $\widetilde{W}(A^{(1)}_1)$ 
is generated by $T=\pi r_1$. 
Its explicit actions $T(x_i)={\overline x_i}$
etc. are determined as the nontrivial solution of 
\be
{\overline x_j}{\overline y_j}=x_j y_j, \quad
{\overline x_{j}}+{\overline y_{j+1}}=x_{j+1}+y_{j}, \quad
\ee
where $x_j=x_{1j}$ and $y_j=x_{2j}$  ($j\in{\bf Z}/n{\bf Z}$). 
This system is a version of the {\em discrete Toda equation}.\cite{H,HT} 
The explicit time evolution is given by
\be \label{eq:Txy}
\overline{x}_j=x_j\dfrac{P_{j-1}}{P_j},\quad 
\overline{y}_j=y_j\dfrac{P_{j}}{P_{j-1}},
\ee
where 
\be
P_{j}=\sum_{a=1}^{n} \overbrace{y_{j+1}\cdots y_{j+a-1}}^{a-1}\,
\overbrace{x_{j+a+1}\cdots x_{j+n}}^{n-a}. 
\ee
The Weyl group $\widetilde{W}(A^{(1)}_{n-1})$ commutes with
this discrete evolution $T$. The explicit actions are
\be\label{eq:sxy}
\begin{array}{ll}
\displaystyle
s_j(x_j)=x_{j+1}\frac{x_j+y_{j+1}}{x_{j+1}+y_j}, 
&\displaystyle 
s_j(y_j)=y_{j+1}\frac{x_{j+1}+y_j}{x_j+y_{j+1}}, \\[3mm]
\displaystyle
s_j(x_{j+1})=x_{j}\frac{x_{j+1}+y_j}{x_j+y_{j+1}}, 
&\displaystyle
s_j(y_{j+1})=y_{j}\frac{x_j+y_{j+1}}{x_{j+1}+y_j},
\end{array}
\ee
$s_j(x_k)=x_k$, $s_j(y_k)=y_k$, ($k \neq j,j+1$).
} 
\end{example}

\begin{example}\quad {\rm 
When $m=3$, we use the variables 
$x_j=x_{1j}$, $y_j=x_{2j}$ and $z_j=x_{3j}$. 
Then $\pi(x_j)=y_j$, $\pi(y_j)=z_j$, $\pi(z_j)=x_j$,
$\rho(x_j)=x_{j+1}$, $\rho(y_j)=y_{j+1}$ and
$\rho(z_j)=z_{j+1}$ ($j\in{\bf Z}/n{\bf Z}$).
The translation subgroup of 
$\widetilde{W}(A^{(1)}_2)$ is generated by 
\be
T_1=\pi r_2 r_1,\quad T_2=r_1 \pi r_2.
\ee
We describe the action of $T_1$, then the action of 
$T_2=\pi T_1 \pi^{-1}$ is given by the rotation.
The actions $T(x_i)={\overline x_i}$ etc. are determined from the 
system of algebraic equations 
\def\ox{{\overline x}}
\def\oy{{\overline y}}
\def\oz{{\overline z}}
\be\label{eq:Txyz}
\begin{array}{l}\smallskip
\ox_j\oy_j\oz_j=x_j y_j z_j,\\ \smallskip
\ox_j\oy_j+\ox_j\oz_{j+1}+\oy_{j+1}\oz_{j+1}=
y_jz_j+y_j x_{j+1}+ z_{j+1} x_{j+1},\\
\ox_j+\oy_{j+1}+\oz_{j+2}=y_j+z_{j+1}+x_{j+2}. 
\end{array}
\ee
as a unique rational solution such that 
\be
\ox_1\cdots \ox_n=y_1\cdots y_n,\quad 
\oy_1\cdots \oy_n=z_1\cdots z_n,\quad 
\oz_1\cdots \oz_n=x_1\cdots x_n. 
\ee
The system (\ref{eq:Txyz}) 
can be regarded as a generalization of the discrete
Toda equation. 
The explicit formulas 
for the time evolution $T_1$
are given in the form
\be
\ox_j=x_j \,\dfrac{F_{j-1}}{F_j},\quad
\oy_j=y_j \,\dfrac{G_{j-1}F_{j}}{G_jF_{j-1}},\quad
\oz_j=z_j \,\dfrac{G_{j}}{G_{j-1}}. 
\ee
Here $F_j$ and $G_j$ are polynomials 
in $x$, $y$, $z$ such that
$F_j=\rho^j(F_0)$, $G_j=\rho^j(G_0)$. 
When $n=3$ for example, $F_0$ and $G_0$ are 
given explicitly as
\be
\begin{array}{l}
F_0=x_1x_2{x_3}^2+x_1x_2x_3y_1+x_2x_3y_1y_2+x_3y_1y_2z_1+x_1x_2x_3z_2\\ 
\qquad\ +\ y_1y_2z_1z_2+x_2x_3y_1z_3+x_3y_1z_1z_3+x_2x_3z_2z_3, \\
G_0=x_2x_3+x_3z_1+z_1z_2. 
\end{array}
\ee
We remark that the affine Weyl group 
$\widetilde{W}(A^{(1)}_{n-1})$ provides 
B\"acklund transformations for the 
time evolutions $T_1$, $T_2$. 
}
\end{example}

Our dynamical system has the following conserved quantities.  
\begin{prop}
The characteristic polynomial of the matrix 
\be
M=A(x_1)\cdots A(x_n)
\ee
is invariant under the actions of the affine Weyl group 
$\widetilde{W}(A^{(1)}_{m-1}) \times \widetilde{W}(A^{(1)}_{n-1})$.
\end{prop}

\noindent
In fact, 
by Lemma \ref{lem:soln}, 
the matrix $M$ itself is invariant under the action of 
$s_i$.  On the other hand, the action of each 
$r_j(M)$ is conjugate to $M$ as 
we have seen in Lemma \ref{lem:GM}.

\begin{example} \quad {\rm 
For $(m,n)=(2,3)$, 
we have the following invariants.\footnote{
By the $m \leftrightarrow n$ duality, the same invariants can also be
obtained as the characteristic polynomial of a $2 \times 2$ matrix.}
\be
\begin{array}l
\det[A(x)A(y)+w I]=
w^3+z^2+w^2(x_1y_1+x_2y_2+x_3y_3)+\\[2mm]
z(x_1x_2x_3+y_1y_2y_3)+w(x_1x_2y_1y_2+x_2x_3y_2y_3+x_3x_1y_3y_1)+\\[2mm]
zw(x_1+x_2+x_3+y_1+y_2+y_3)+x_1x_2x_3y_1y_2y_3.
\end{array}
\ee
}
\end{example}

Since the Weyl group actions are {\em tropical} \cite{K}
(i.e. subtraction-free birational mappings), 
there exists a combinatorial counterpart
obtained by the ultra-discretization \cite{ultra}
\be
a\times b \to a+b,\quad a+b \to \min(a,b). 
\ee
Corresponding combinatorial dynamical system can be viewed
as a periodic version of the Box-Ball systems.
Similarly to the original Box-Ball system \cite{TS} and
their generalizations \cite{bbsgen}, this dynamical system also 
has soliton type solutions.

\par\medskip
The transformations as in Example \ref{ex:m2} are
quite similar to those for the $q$-Painlev\'e equation
$q$-$P_{IV}$ and their B\"acklund transformations. 
In fact, the realization of the Weyl group 
as in (\ref{eq:GM}), (\ref{eq:GM2}) is defined exactly in 
the same way as that of 
the B\"acklund transformations of the Painlev\'e equations 
\cite{NY}
and their Lie-theoretic generalization 
studied in \cite{NY2}, Section 4. 
The $q$-$P_{IV}$ can be viewed as a non-autonomous 
deformation of
our integrable system for $(m,n)=(2,3)$ of this paper. 

In order to see this surprising coincidence more precisely, 
let us reformulate Example \ref{ex:m2}.  
Putting $x_j y_j=b_j$, $a_j=b_j/b_{j+1}$
and $x_j x_{j+1}=\varphi_j b_j$, we have
\be 
{\overline b_j}=b_j, \quad
\frac{\varphi_j+1}{\overline{\varphi_j}+a_j}=\frac{x_{j+1}}{\overline{x_j}}.
\ee
Eliminating the $x_j$ variables, one obtain
\be
(1+\varphi_j)(1+\frac{1}{\varphi_{j+1}})=
(1+\frac{\overline{\varphi_{j+1}}}{a_{j+1}})
(1+\frac{a_j}{\overline{\varphi_j}}).
\ee 
In terms of the variables $a_j$ and $\varphi_j$, 
the time evolution (\ref{eq:Txy}) is rewritten in the form 
\be\label{eq:qPeq}
\begin{array}l
\displaystyle
{\overline a_j}=a_j, \quad
{\overline \varphi_j}=a_j \varphi_{j+1} \frac{g_{j+2}}{g_j}, \\[4mm]
g_j=1+\varphi_j+\varphi_j \varphi_{j+1}+\cdots+
\varphi_j\cdots \varphi_{j+n-2}.
\end{array}
\ee
The Weyl group $\widetilde{W}(A^{(1)}_{n-1})$ commutes with
this discrete evolution. 
In terms of variables $a_j$ and $\varphi_j$, the B\"acklund 
transformations (\ref{eq:sxy}) are given by
\be\label{eq:qPsym}
\begin{array}{lll}
\displaystyle s_j(a_j)=\frac{1}{a_j},
&\displaystyle s_j({a_{j+1}})=a_{j+1}a_j,
&\displaystyle s_j({a_{i-1}})=a_{i-1}a_j,\\
\displaystyle s_j(\varphi_j)=\frac{\varphi_j}{a_j},
&\displaystyle s_j(\varphi_{j+1})=
\varphi_{j+1}\frac{a_j+\varphi_j}{1+\varphi_j},
&\displaystyle s_j(\varphi_{i-1})=
\varphi_{i-1}\frac{1+\varphi_j}{1+\varphi_j/a_j},\\
\end{array}
\ee
and $s_j({a_k})=a_k$, $s_j(\varphi_k)=\varphi_k$ 
for $k \neq j,j\pm 1$ (mod $n$).

In this setting, the parameters $a_j$ are subject to the constraint
\be
q \equiv a_0a_1\cdots a_{n-1}=1.
\ee
However, the Weyl group representation in (\ref{eq:qPsym}) and
its commutativity with the evolution (\ref{eq:qPeq}) 
survive also for $q \neq 1$. In this non-autonomous situation,
the equation (\ref{eq:qPeq}) gives a generalized $q$-Painlev\'e
equation which has the $\widetilde{W}(A^{(1)}_{n-1})$ 
B\"acklund transformations (\ref{eq:qPsym}); 
the case $n=3$ recovers the $q$-$P_{IV}$ equation 
\cite{KNY}.  
The non-autonomous versions of 
the cases with 
$m \geq 3$ could be considered as a 
$q$-Painlev\'e hierarchy 
with multi-discrete times. 


\end{document}